\documentclass[%
 reprint,
 amsmath,amssymb,
 aps, prl,
]{revtex4-1}

\usepackage{graphicx}	% Including figure files
\usepackage{amsmath}	% Advanced maths commands
\usepackage{amssymb}	% Extra maths symbols

\usepackage[english]{babel}
\usepackage{graphicx}
\usepackage{latexsym}
\usepackage{floatflt}
\usepackage{float}
\usepackage{units}
\usepackage{hyperref}
\usepackage{amsmath}
\usepackage{gensymb}
\usepackage{mathdots}
\usepackage{color}
\usepackage{xcolor}
\usepackage{mathtools}
\usepackage{lipsum}

\usepackage{dcolumn}% Align table columns on decimal point
\usepackage{bm}% bold math

\newcommand{\id}{{\rm d}}

\begin{document}

\preprint{APS/123-QED}

\title{A cosmic string solution to the radio synchrotron background}

\author{Bryce Cyr}
\email{bryce.cyr@manchester.ac.uk}
\author{Jens Chluba}
\author{Sandeep Kumar Acharya}

\affiliation{%
Jodrell Bank Centre for Astrophysics, School of Physics and Astronomy, The University of Manchester, Manchester M13 9PL, U.K.
}%

\begin{abstract}
We investigate the low-frequency spectral emission from a network of superconducting cosmic string loops in hopes of explaining the observed radio synchrotron background. After considering constraints from a variety of astrophysical and cosmological measurements, we identify a best-fit solution with string tension $G\mu \simeq 6.5 \times 10^{-12}$ and current $\mathcal{I} \simeq 2.5 \times 10^6$ GeV. 
%This model has a $\Delta \chi^2_{\rm r} \simeq 0.68$ when compared with a simple power-law model, yielding a convincing fit to the data. 
%
This model yields a convincing fit to the data and may be testable in the near future by spectral distortion (TMS, BISOU) and 21 cm experiments (HERA, SKA, REACH).
We also find that soft photon heating protects us against current constraints from global $21$ cm experiments. 
\end{abstract}

\maketitle

%%%%%%%%%%%%%%%%%%%%%%%%%%%%%%%%%%%%%%%%%%%%%%%%%%%%%%%%%%%%%%%%
\section{\label{sec:level1}Introduction}
%%%%%%%%%%%%%%%%%%%%%%%%%%%%%%%%%%%%%%%%%%%%%%%%%%%%%%%%%%%%%%%%
Over the past decade, radio frequency observations of the sky have received much attention. This has been spurred by two anomalous signals that remain unexplained within 
%the current paradigm. 
$\Lambda$CDM.
%count -2
First, the observations of short-duration extragalactic radio pulses dubbed ``fast radio bursts" still lack a proven progenitor model \cite{Cordes2019}, though both astrophysical \cite{Zhang2020} and more exotic models \cite{Yu2014, Ye2017, Imtiaz2020} have been proposed. 
%
%Secondly, the detection of an excess radio background at frequencies $\nu \lesssim 10$ GHz was originally reported by the ARCADE-2 experiment \cite{Fixsen2009} over a decade ago. Since then, additional low-frequency data has been analyzed, corroborating and extending the radio excess down to $\nu \simeq 20$~MHz \cite{Dowell2018}. This signal is sometimes referred to as the radio synchrotron background (RSB), and its origin is currently unknown \citep[see][for recent reviews]{Singal2018, Singal2022},
%count -68
Second, the detection of an excess radio background at frequencies $\nu \lesssim 10$ GHz \citep{Fixsen2009, Dowell2018}, often referred to as the {\it radio synchrotron background} (RSB), still remains unexplained \citep[see][for recent reviews]{Singal2018, Singal2022}.
%count +31
%
%The ARCADE-2 collaboration \cite{Fixsen2009}, as well as \citet{Dowell2018} both find that power laws offer a convincing fit to the RSB data 
%count: -21
%%%%%%
%\footnote{Although the spectral index found by these two groups agrees, Dowell and Taylor find a slightly higher normalization at a reference frequency of $\nu = 310$~MHz.}. 
%However, they make no attempt to subtract off any contribution from resolved, discrete extragalactic sources, nor give a physical explanation for the RSB. A reasonable fit to this minimal extragalactic background from discrete sources is given by $T_{\rm bg}(\nu) \simeq 0.23 \, {\rm K} \, (\nu/{\rm GHz})^{-2.7}$ \cite{Gervasi2008}. 
%count: -62 words
%
%Taking this into account, a best-fit power law to the RSB is 
%count -12
%
Including the minimal extragalactic background from discrete sources \cite{Gervasi2008}, a power-law representation of the RSB is given by
%count +17
%%%%%%%%%%%%%%%%%%%%%%%%%%%%%%%%%%%%%%%%%%%%%%%%%%
\begin{equation} \label{eq:PowLaw}
T_{\rm RSB}(\nu) \simeq 1.230 \, {\rm K} \left( \frac{\nu}{{\rm GHz}}\right)^{-2.555}.
\end{equation}
%%%%%%%%%%%%%%%%%%%%%%%%%%%%%%%%%%%%%%%%%%%%%%%%%%
The background surface brightness from known classes of extragalactic point sources appears to be at least a factor of 3 smaller than the observed radio excess \citep{Gervasi2008}, with possible explanations having galactic or extragalactic origins \citep{SC2013}. Faint, unresolved/undetected point sources could in principle account for the excess; however, in this case the signal is expected to be anisotropic \citep{H2014}. Current measurements regarding the isotropy of the signal indicate that it may have significantly more angular power on sub-degree scales at MHz frequencies than would be expected from simple point source models \cite{OSHHL2022, Cowie2023}. Other exotic explanations of the radio excess include synchrotron emission from high energy particles \citep{FLRT2011}, emission from accreting stellar black holes \citep{ECLDSM2018}, primordial black holes \citep{MK2021}, 
%and dark matter-dark photon interactions \citep{Caputo2022}. 
%count -6
and others \citep{Caputo2022}.
%count +3
%However, all of these models have their limitations: Black holes emit radio photons as well as energetic photons which can ionize the universe, leading to strong CMB anisotropy constraints \citep{Acharya2022}. Similarly, exotic particle physics models involving dark photons require significant fine-tuning, and are likely not realized in our universe \citep{AC2022}. 
% count -50
However, each of these models have their limitations \citep{Acharya2022, AC2022},
% count +10
%In this work, 
and here
%count -1
we propose a model of superconducting cosmic strings that could explain the anomalous RSB, while being consistent with cosmological and astrophysical datasets.
%count -1

Cosmic strings are a class of topological defect that may form at the interface of cosmological phase transitions in the very early universe, provided that the true vacuum manifold of the considered theory is both degenerate and simply connected. If such a phase transition occurs, the Kibble mechanism ensures that a stable network of long strings forms, following a scaling distribution as the universe expands \cite{Vilenkin2000,Kibble1980, Kibble1976, Magueijo2000}. Further analysis of these models indicates that an abundant distribution of smaller, sub-Hubble string loops may also be sourced through the intersections and self-intersections of long strings in the network. Since these original studies, Nambu-Goto simulations (which neglect the finite thickness of the strings) have been performed, demonstrating the existence \footnote{A separate set of field theory simulations which attempt to resolve the cosmic string cores have been performed and do not see a long-lived loop distribution \cite{Hindmarsh2008,Hindmarsh2017}. We note that there are significant numerical challenges involved when simulating such a wide range of scales from the string width to the Hubble radius, and utilize the loop distributions given from Nambu-Goto simulations in our work.} of this loop distribution \cite{Vanchurin2005, Martins2005, Ringeval2005,Lorenz2010,Blanco-Pillado2013}. These loops 
%have the potential to
%count -4
can
%count +1
give rise to a plethora of astrophysical and cosmological signatures, from a stochastic gravitational wave background \cite{Vachaspati1984, Ellis2020, Blanco-Pillado2021}, to the seeds of massive black holes \cite{Bramberger2015, Brandenberger2021, Cyr2022}, and more. 
%Importantly, 
%count -1
Gravitational signatures of cosmic strings are typically stronger for phase transitions that take place at 
%a
%count -1
higher energy scales. Therefore, cosmic strings represent a well motivated class of models which help 
%to
%count -1
probe aspects of particle physics not accessible by conventional collider searches. Their detection (or non-detection) provides us with invaluable glimpses into the nature of the symmetry breaking patterns that the universe may have undergone at the earliest epochs and highest energies. 

It has also been demonstrated that some symmetry breaking patterns can bestow the strings with superconductive properties \cite{Witten1984, Ostriker1986}. In these models, $U(1)_{\rm em}$ may be broken in the core of the string, and significant currents can be generated as the loops oscillate and evolve in a background of magnetic fields. These superconducting loops are capable of emitting strong bursts of electromagnetic radiation \cite{Babul1987, Vilenkin1986}, which has led to a variety of constraints on the model space \cite{Cai2012, Tashiro2012, Tashiro2012d,Miyamoto2012, Brandenberger2019},  most recently in \cite{Cyr2023}. As a simplifying assumption, the current, $\mathcal{I}$, of a superconducting loop distribution is taken to be both time- and loop-length independent. In reality, the current is generated dynamically by local magnetic fields, but simulations of this have not been performed and represent an important avenue for future study. As a result, superconducting models are
%completely
%count -1
distinguished by two independent parameters, 
%The gravitational effects are set by the string tension, $G\mu$ ($G$ here is the gravitational constant), while electromagnetic phenomena are typically controlled by $\mathcal{I}$. 
%count -24
the loop current, $\mathcal{I}$, and string tension, $G\mu$ ($G$ here is the gravitational constant).
%count +14

A background of radio photons can be sourced from the incoherent electromagnetic bursts of a network of superconducting string loops. In recent work \citep{Cyr2023}, we considered a variety of constraints on superconducting strings arising from this spectral emission. In particular, we considered the data from ARCADE-2 and other low-frequency experiments measuring the radio background as strict upper limits on the amount of emission allowed by the loop distribution. In this \textit{Letter}, we now examine the viability of a %possible 
%count -1
cosmic string explanation to the observed RSB, showing that a small region in parameter space 
%located around $G\mu \simeq 6.5 \times 10^{-12}$ and $\mathcal{I} \simeq 2.5 \times 10^6$ GeV comes convincingly close. 
%count -17
provides a convincing solution.
%count +4

In the next section, we discuss the mechanism for photon emission from superconducting cosmic strings. Afterwards, we briefly review current constraints and provide details about the model that fits
%very
%count -1
well with current observations \citep[see][for more details]{Cyr2023}. Finally, we discuss 
%the
%count -1
possible implications and avenues for further study before concluding. Except where stated, we use natural units with $\hbar = c = k = 1$.

%%%%%%%%%%%%%%%%%%%%%%%%%%%%%%%%%%%%%%%%%%%%%%%%%%%%%%%%%%%%%%%%
\vspace{-3mm}
\section{Photon production mechanism}
\label{sec:level2} 
\vspace{-3mm}
%%%%%%%%%%%%%%%%%%%%%%%%%%%%%%%%%%%%%%%%%%%%%%%%%%%%%%%%%%%%%%%%
At a given initial time, $t_{\rm i}$, simulations indicate that most string loops are formed with a length given by some fraction of the Hubble scale,  $L_{\rm i} \simeq \beta t_{\rm i}$ where $\beta \simeq \mathcal{O}(0.1)$. Upon formation, the loops undergo oscillations (with period $T \simeq L$), which leads to the formation of substructure on the strings known as cusps and kinks. The substructures decay rapidly once formed, leading to violent bursts of electromagnetic radiation over a wide spectral range if the strings are superconducting. Cusp annihilations typically emit the most energy, and so we neglect the effects of kinks in what follows. 

The oscillation averaged power emitted by a single cusp annihilation is given by $P_{\gamma} \simeq \Gamma_{\gamma} \mathcal{I} \mu^{1/2}$, where $\Gamma_{\gamma} \simeq \mathcal{O}(10)$ is determined by simulations \cite{Cai2012, Vilenkin1986}. In contrast, the energy carried away by gravitational waves is $P_{\rm g} \simeq \Gamma_{\rm g} G\mu^2$, with $\Gamma_{\rm g} \simeq \mathcal{O}(100)$ \cite{Vachaspati1984}.
%From this it follows that
%count -5
Thus,
%count +1
loops shrink as they emit energy into gravitational and electromagnetic radiation. 
%One can find that 
%count -4
At any time after formation ($t_{\rm i}$), the loop size is given by
%count +2
%%%%%%%%%%%%%%%%%%%%%%%%%%%%%%%%%%%%%%%%%%%%%%%%%%
\begin{equation}
L(t) = L_{\rm i}(1+\lambda) - \Gamma G\mu t,
\end{equation}
%%%%%%%%%%%%%%%%%%%%%%%%%%%%%%%%%%%%%%%%%%%%%%%%%%
where $\Gamma G\mu = (P_{\rm g} + P_{\rm \gamma})/\mu$, and $\lambda = \Gamma G\mu/\beta$ is a measure of how quickly after formation a loop will evaporate. Loops with $\lambda \geq 1/\beta$ evaporate within one oscillation time, which can lead to a breakdown of the cusp annihilation formalism described here. 

Sufficiently long after the phase transition which formed the strings, a distribution of loops will be established on scales with $L \lesssim \beta t$. The number density of loops per unit length in the matter-dominated era is then given by \citep{Blanco-Pillado2013,Cyr2023}
%%%%%%%%%%%%%%%%%%%%%%%%%%%%%%%%%%%%%%%%%%%%%%%%%%
\begin{equation} \label{eq:loopDist}
\frac{\id N_{\rm{loops}}}{\id L}\approx\frac{\alpha\left( 1 + \lambda\right)^{3/2} t_{\rm eq}^{1/2}}{t^{2}(L+\Gamma G\mu t)^{5/2}}+  \frac{\alpha_{\rm m}\left( 1 + \lambda\right)}{t^{2}(L+\Gamma G\mu t)^{2}}.
\end{equation}
%%%%%%%%%%%%%%%%%%%%%%%%%%%%%%%%%%%%%%%%%%%%%%%%%%
%%%%%%%%%%%%%%%%%%%%%%%%%%%%%%%%%%%%%%%%%%%%%%%%%%
\begin{figure}
\includegraphics[width=\columnwidth]{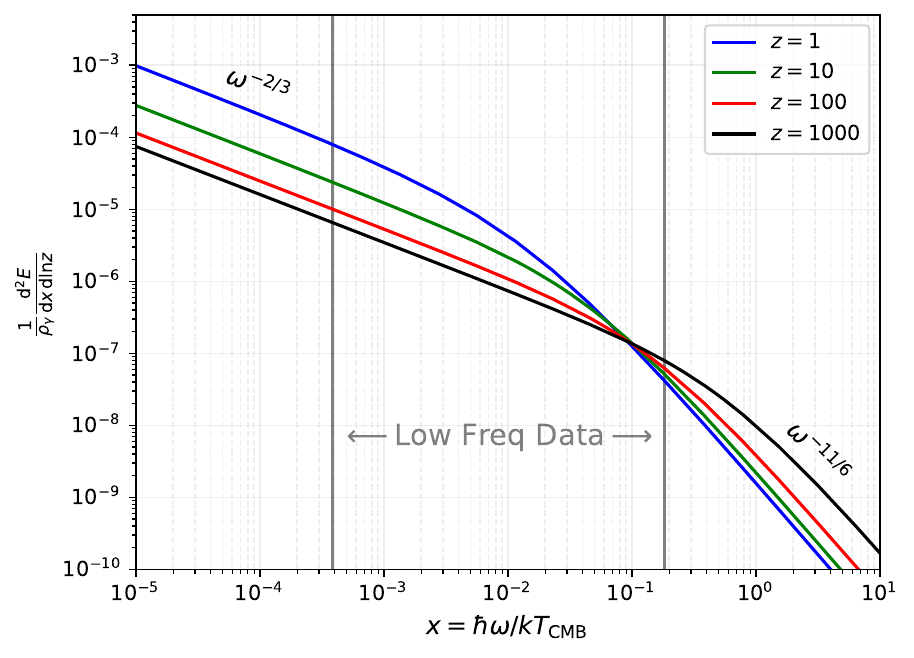}
\caption{The instantaneous emission spectrum from a distribution of string loops with $G\mu \simeq 6.5 \times 10^{-12}$ and $\mathcal{I} \simeq 2.5 \times 10^6$~GeV, normalized to the CMB energy density at each respective redshift. The knee position slowly shifts to lower frequencies at later times. The radio spectrum observed comes from the integrated effect of emission from all loops at all redshifts between the last scattering surface and today. 
%
%The frequency coverage analyzed by ARCADE-2 \cite{Fixsen2009} as well as \citet{Dowell2018} is indicated by the vertical lines.
%count -17
}
\label{fig:BestFitSpectrum}
\end{figure}
%%%%%%%%%%%%%%%%%%%%%%%%%%%%%%%%%%%%%%%%%%%%%%%%%%
%%%%%%%%%%%%%%%%%%%%%%%%%%%%%%%%%%%%%%%%%%%%%%%%%%
\begin{figure*}
\centering 
\includegraphics[width=\columnwidth]{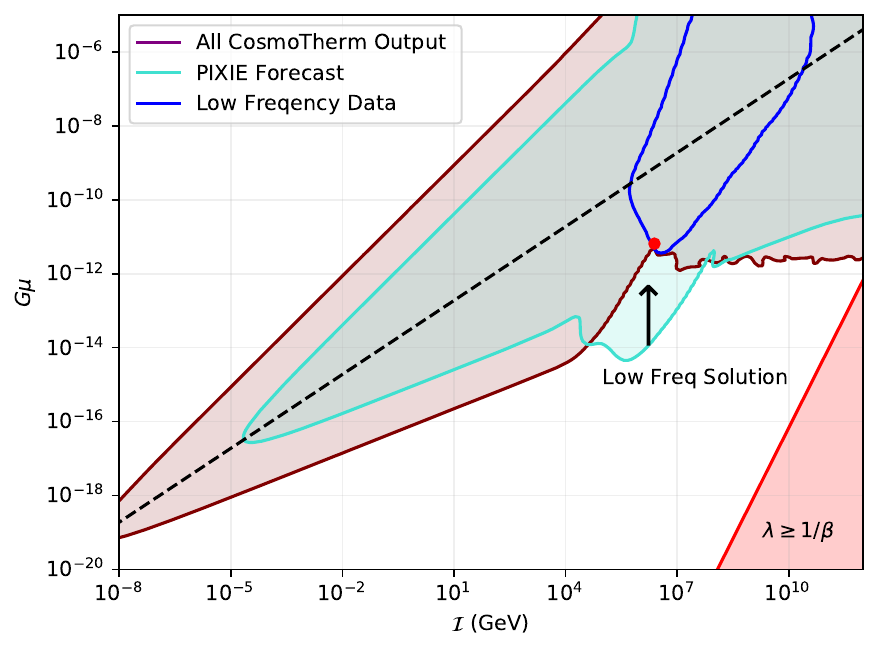}
\hspace{4mm}
\includegraphics[width=\columnwidth]{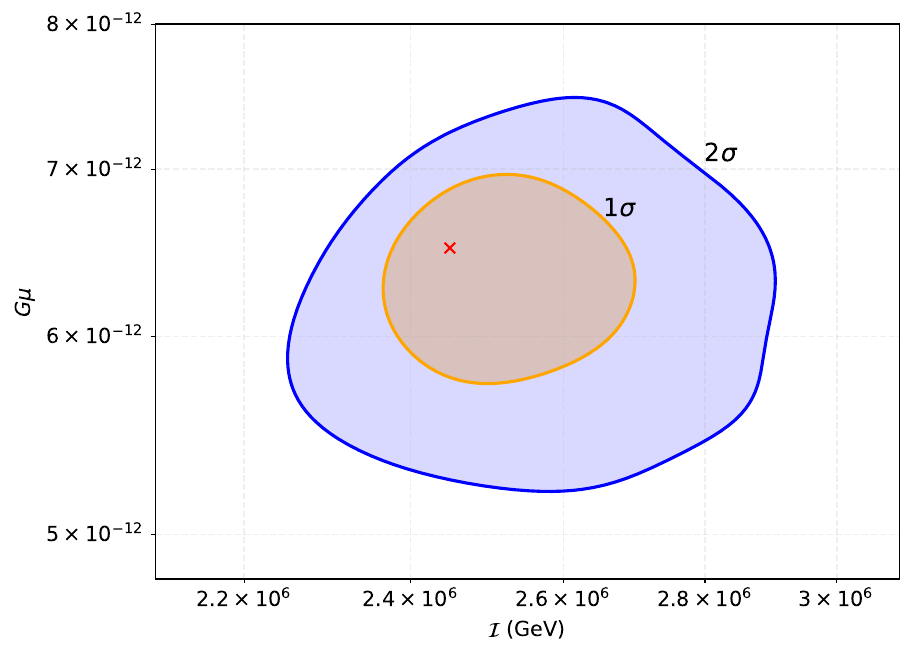}
%\caption{Left: $2\sigma$ constraints from a variety of astrophysical and cosmological observations \cite{Cyr2023}. Above the dashed line, gravitational effects determine the lifetime of a loop ($P_{\rm g} \geq P_{\gamma}$), while below, electromagnetic injection dominates. The red region with $\lambda \geq 1/\beta$ represents where the oscillation-averaged cusp annihilation formalism breaks down. Right: The combined likelihood analysis with forecasted constraints from a PIXIE-type instrument \cite{Kogut2011PIXIE}. The position of the best fitting string solution to the RSB is marked in both plots by the red dot. }
\caption{Left: $2\sigma$ constraints from a combined likelihood analysis of a variety of 
%astrophysical and cosmological
%count -3
observations \cite{Cyr2023}. Above the dashed line, gravitational effects determine the lifetime of a loop, 
%($P_{\rm g} \geq P_{\gamma}$), 
%count -2
while below electromagnetic injection dominates. The red region with $\lambda \geq 1/\beta$ indicates where the oscillation-averaged cusp annihilation formalism breaks down. 
%The turquoise region is the forecasted constraints from a PIXIE-type instrument \cite{Kogut2011PIXIE}. 
%count -12
Right: $1\sigma$ (orange) and $2\sigma$ (blue) regions around our best fit solution of $G\mu \simeq 6.5 \times 10^{-12}$ and $\mathcal{I} \simeq 2.5 \times 10^6$ GeV (marked by the red x).}
\label{fig:AllConstraints}
\end{figure*}
%%%%%%%%%%%%%%%%%%%%%%%%%%%%%%%%%%%%%%%%%%%%%%%%%%
Here, $\alpha = 0.18$ and $\alpha_{\rm m} = \alpha / \sqrt{\beta}$. The first term represents loops forming in the radiation era, while the second is from the larger loops forming during matter domination. A similar form can be found for the evolution in the radiation era, but these are irrelevant for the production of a radio background from loops. 

%\begin{eqnarray}
%\frac{\id N_{\rm{loops}}}{\id L} = \begin{dcases} &\frac{\alpha\left( 1 + \lambda\right)^{3/2}}{t^{3/2}(L+\Gamma G\mu t)^{5/2}} 
%\\
% &\frac{\alpha\left( 1 + \lambda\right)^{3/2} t_{\rm eq}^{1/2}}{t^{2}(L+\Gamma G\mu t)^{5/2}}+  \frac{\alpha_{\rm m}\left( 1 + \lambda\right)}{t^{2}%(L+\Gamma G\mu t)^{2}},
%\end{dcases}
%\end{eqnarray}

It is evident from Eq.~\eqref{eq:loopDist} that $L_{\rm c}(t) = \Gamma G\mu t$ defines a special length scale. Indeed, one can show that loops with $L \leq L_{\rm c}(t)$ decay within one Hubble time, so $L_{\rm c}$ acts as the cutoff between a more transient population of decaying loops, and the longer-lived set with $L \geq L_{\rm c}(t)$.

The oscillation-averaged photon spectrum produced by a single cusp event (per unit time) is given by \citep{Cai2012, Cyr2023}
%%%%%%%%%%%%%%%%%%%%%%%%%%%%%%%%%%%%%%%%%%%%%%%%%%
\begin{equation} \label{eq:cuspSpectrum}
\frac{\id^2 E^{\rm c}_{\gamma}}{\id \omega \id t} \simeq \left( \frac{\Gamma_{\gamma}}{3}\right) \frac{\mathcal{I}^2 L^{1/3}}{\omega^{2/3}}.
\end{equation}
%%%%%%%%%%%%%%%%%%%%%%%%%%%%%%%%%%%%%%%%%%%%%%%%%%
%The spectrum has both high- and low-frequency cutoffs. The high-frequency cutoff has been estimated as the energy at which the produced photons begin to exceed the energy budget allowed by a cusp annihilation, $\omega_{\rm max} \simeq \mu^{3/2}/\mathcal{I}^3 L$, while the low-frequency cutoff is set by the frequency below which produced photons are rapidly absorbed by free-free processes in the background ($\omega_{\rm c}$). 
%count -64
The spectrum has a high-frequency cutoff at  $\omega_{\rm max} \simeq \mu^{3/2}/\mathcal{I}^3 L$, where the produced photons begin to exceed the energy budget allowed by a cusp annihilation.
%count +31

%On average, each loop emits photons into the background with a spectrum given by Eq.~\eqref{eq:cuspSpectrum} between frequencies $\omega_{\rm c} \leq \omega \leq \omega_{\rm max}$. 
% count -25
Using Eq.~\eqref{eq:cuspSpectrum},
%count + 2
the full emission spectrum from the loop distribution can be obtained by the appropriately weighted integral,
%%%%%%%%%%%%%%%%%%%%%%%%%%%%%%%%%%%%%%%%%%%%%%%%%%
\begin{equation} \label{eq:fullSpectrum}
\frac{\id^2 E_{\gamma}}{\id \omega \id t} = \int_0^{L_{\rm max}} \id L \frac{\id^2 E^{\rm c}_{\gamma}}{\id \omega \id t} \frac{\id N_{\rm loops}}{\id L},
\end{equation}
%%%%%%%%%%%%%%%%%%%%%%%%%%%%%%%%%%%%%%%%%%%%%%%%%%
with $L_{\rm max} = \mu^{3/2}/\mathcal{I}^3 \omega$, encoding the fact that only small loops can give rise to arbitrarily high energy photons 
\citep[see][for details]{Cyr2023}. 
%count +4
%The full emission spectrum thus has no upper frequency cutoff, but still maintains a low-frequency cutoff at $\omega_{\rm c}$. 
%count -19
%For a full analysis of the spectral properties, as well as useful asymptotic expressions, we refer the reader to our recent work \cite{Cyr2023}.
%count -23

In Fig.~\ref{fig:BestFitSpectrum}, we illustrate the evolution of this full spectrum over a range of redshifts for our best-fit string parameters %which we discuss below. 
discussed below.
%count -2
The spectrum is well-fit by a broken power law, going as $\omega^{-2/3}$ at low frequencies, and as $\omega^{-11/6}$ in the ultraviolet. The position of the knee is determined by the $\omega_{\rm max}$ produced by loops at the cutoff length $L_{\rm c}$. The dropoff in the high frequency spectrum comes from the fact that these photons are only produced by the short-lived population of loops with $L \lesssim L_{\rm c}(t)$.

%%%%%%%%%%%%%%%%%%%%%%%%%%%%%%%%%%%%%%%%%%%%%%%%%%%%%%%%%%%%%%%%
\section{\label{sec:level3} A radio synchrotron background from superconducting cosmic strings}
%%%%%%%%%%%%%%%%%%%%%%%%%%%%%%%%%%%%%%%%%%%%%%%%%%%%%%%%%%%%%%%%
In our recent work \cite{Cyr2023}, we considered the constraints on $G\mu$ and $\mathcal{I}$ 
%that could be
%count -3
inferred from the effects of spectral emission by loops on a variety of astrophysical and cosmological phenomena. Specifically, we derived constraints from CMB anisotropies, the optical depth to reionization \cite{Planck2018paramsShort}, and spectral distortion data \cite{Fixsen1996}.
We furthermore used the radio background data \cite{Fixsen2009, Dowell2018} and the EDGES observation \cite{Edges2018} as upper limits on the model space.
In addition, we forecasted constraints from $\mu$, as well as non-$\mu$, non-$y$ type distortions envisioning a PIXIE-type spectral distortion experiment \cite{Kogut2011PIXIE, Kogut2016SPIE}. 
A summary of these constraints and forecasts (at $2\sigma$) can be found in Fig.~\ref{fig:AllConstraints}, which highlights that a wide range in parameter space can already be excluded.
%%%%%%%%%%%%%%%%%%%%%%%%%%%%%%%%%%%%%%%%%%%%%%%%%%
\begin{figure*}
\centering 
\includegraphics[width=\columnwidth]{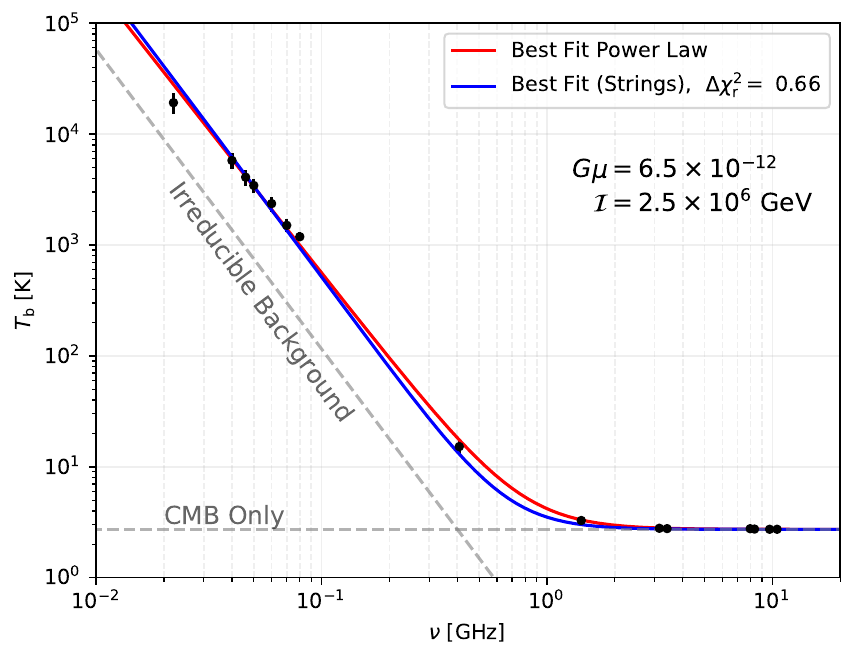}
\hspace{4mm}
\includegraphics[width=\columnwidth]{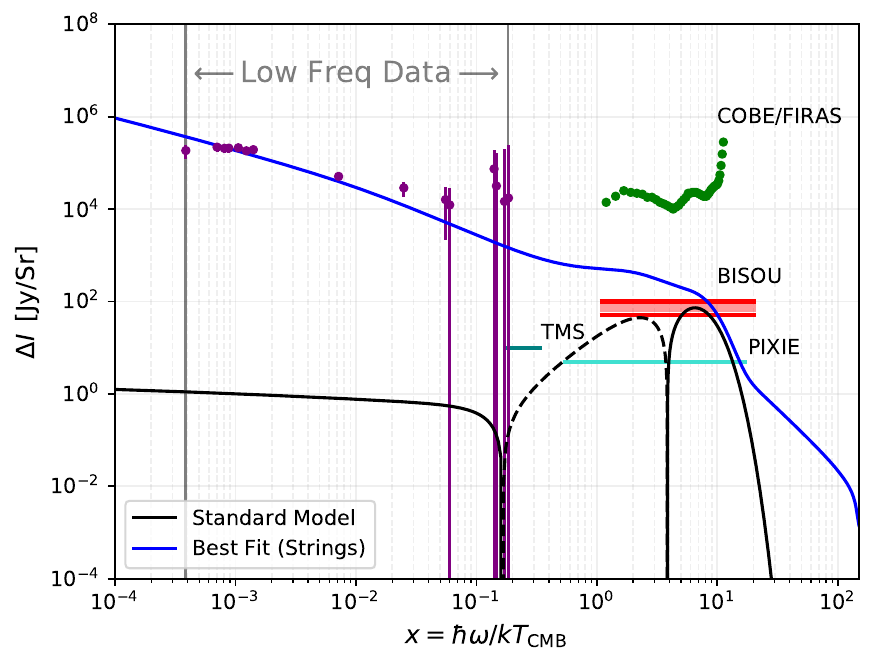}
\caption{Left: A comparison of the low frequency brightness temperature ($T_{\rm b} = c^2 I /2 k \nu^2$) from a best fit power law (red) and the cosmic string solution (blue). 
%We also show the observations by ARCADE-2 and other low-frequency experiments, as well as the contribution from known classes of extragalactic point sources \cite{Fixsen2009, Dowell2018, Gervasi2008}. 
%count -26
Cosmic strings offer a convincing fit to the radio data \cite{Fixsen2009, Dowell2018, Gervasi2008}.
%count +2
Right: the raw spectral distortion output from \texttt{CosmoTherm} at $z = 0$. The black line shows the generation of distortions with no strings (reionization effects drive the standard model signal), while the blue line is our string model. The forecasted sensitivities for various experiments
%curves for BISOU \cite{BISOU}, TMS \cite{Jose2020TMS}, and a PIXIE-like experiment \cite{Kogut2011PIXIE, Kogut2016SPIE} are shown, indicating 
%count -15
indicate 
%count +1
that this model can soon be probed at high significance.}
\label{fig:LowFreqFits}
\end{figure*}
%%%%%%%%%%%%%%%%%%%%%%%%%%%%%%%%%%%%%%%%%%%%%%%%%%

To determine these constraints, we utilized the numerical code \texttt{CosmoTherm} \footnote{\url{www.Chluba.de/CosmoTherm}} \citep{Chluba2011therm, Chluba2015GreensII, Bolliet2020, Acharya2022},
evaluating the various likelihoods following \citep{Cyr2023}.
%count +8 
%which takes as input an instantaneous emission spectrum such as the one derived in Eq.~\eqref{eq:fullSpectrum}. 
%count -16 
%With this, \texttt{CosmoTherm} solves the thermalization equations for the photons and electrons, computing the resultant spectral distortion at many finely spaced redshift steps. 
%count -23
%\texttt{CosmoTherm} also has modules that allow for the computation of the differential brightness temperature at cosmic dawn ($\delta T_{\rm b}$), and the ionization fractions at any redshift \citep{Acharya2022}. We have also recently added a simple likelihood module \citep{Cyr2023}, which computes the $\Delta\chi^2$ between a model with strings and the null hypothesis (with the exception of the low-frequency data likelihood, which compares against a best-fit power law for the RSB).
%count -69
%
However, this time we instead performed a search for possible cosmic string solutions to the anomalous RSB, 
%In this search, we 
%count -4
identifying a region of parameter space capable of fitting the observed radio data at high significance. This region runs along the left-most edge of the ``low-frequency data" contour in Fig.~\ref{fig:AllConstraints}. Demanding that we do not violate our other constraints, we find that a distribution of superconducting string loops with $G\mu \simeq 6.5 \times 10^{-12}$ and $\mathcal{I} \simeq 2.5 \times 10^6$ GeV produces a low-frequency radio background which offers a very convincing match to the data. 

The left panel of Fig.~\ref{fig:LowFreqFits} shows a comparison between our string solution and the best fit power law we discussed in Eq.~\eqref{eq:PowLaw} ($T_{\rm b} = T_{\rm RSB} + T_{\rm CMB}$). The irreducible background is given by the extragalactic component determined in \cite{Gervasi2008}. We computed the reduced $\Delta \chi^2_{\rm r}$ between an ad-hoc power law fit, and our strings, finding $\Delta \chi^2_{\rm r} \simeq 0.66$, which indicates an impressive fit to the RSB data \cite{Fixsen2009, Dowell2018}. The right panel of Fig.~\ref{fig:LowFreqFits} shows the raw output from \texttt{CosmoTherm} over an expanded bandwidth. Note that \texttt{CosmoTherm} determines the spectral distortion, $\Delta I$, from which the string induced spectrum can be recovered by computing $I_{\rm strings} = I_{\rm BB}(T_{\rm CMB}) + \Delta I$.

We have marked the position of this string solution in both panels of Fig.~\ref{fig:AllConstraints}, and find that it lies near the boundary of the constraints derived from CMB anisotropies. The solution ends up being near a contour in parameter space where the knee frequency of the spectrum (see Fig.~\ref{fig:BestFitSpectrum}) is $\omega_{\rm k} \simeq 13.6$ eV at the time of last scattering. This tells us that ionizing photons are not produced efficiently by the string network at that time, which is why CMB anisotropy constraints relax. 
%so much for $\mathcal{I} \gtrsim 10^4$ GeV. 
%count -4
Without this effect, no RSB solution would be possible. 

Another crucial ingredient for the existence of this solution is the inclusion of a newly discovered effect known as \textit{soft photon heating} (SPH) \cite{Acharya2023}, which describes the interplay between extra low-frequency backgrounds and $21$ cm observables such as the brightness temperature at cosmic dawn, $\delta T_{\rm b}$. 
In \cite{Acharya2023}, it was shown that the presence of sufficiently steep low-frequency backgrounds (spectral index $\gtrsim 2.5$ at $\nu \lesssim 1$ GHz) can have a dramatic effect on the absorption depth of $\delta T_{\rm b}$. 

It is well known that the amplitude of $\delta T_{\rm b}$ is proportional to $T_{\rm rad}/T_{\rm spin}$, where $T_{\rm rad}$ is the brightness temperature of background radiation at the $21$ cm wavelength. During cosmic dawn, the dominant contribution to the spin temperature ($T_{\rm spin}$) comes from the kinetic motion of the hydrogen atoms. In \citet{Acharya2023}, we showed that the presence of additional radio backgrounds causes a significant increase in the spin temperature as the hydrogen atoms are heated up. Therefore, SPH generically dampens the amplitude of $\delta T_{\rm b}$. 

\begin{figure}
\includegraphics[width=\columnwidth]{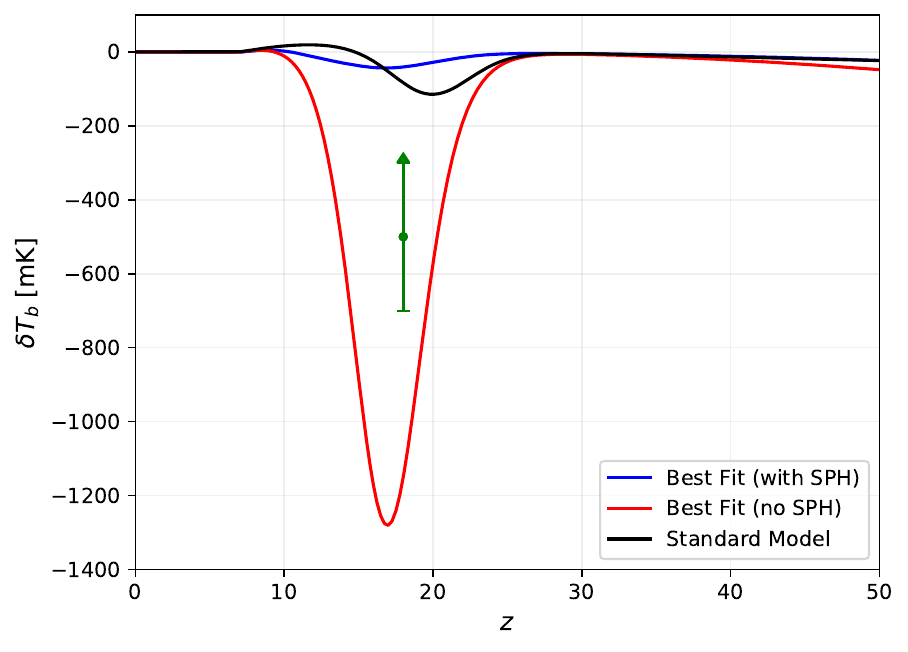}
\caption{The differential brightness temperature at cosmic dawn for our best fit string model with (blue) and without (red) the inclusion of soft photon heating. The EDGES datapoint is marked in green. Without soft photon heating, this model would be ruled out at more than $2\sigma$ by taking EDGES as a strict bound on the amplitude of $\delta T_{\rm b}$.}
\label{fig:dTbFit}
\end{figure}

The EDGES collaboration has claimed the first detection of the differential brightness temperature at cosmic dawn ($z \simeq 17$) \cite{Edges2018}.
%, with $\delta T_{\rm b} = -500$ mK and a $1\sigma$ error of $200$ mK \cite{Edges2018}. While the validity of this detection is in question \cite{Saras2022}, it is still possible to use EDGES as a strict upper bound on the amplitude of the global 21 cm signal. 
%count -45
Before SPH was understood, it was claimed that if the RSB was in place at cosmic dawn, it would produce a $\delta T_{\rm b}$ far in excess of what EDGES observed \cite{Feng2018}. Importantly, SPH can reconcile these two observations. 
This is illustrated in Fig.~\ref{fig:dTbFit}, from which
%count +8
%In Fig.~\ref{fig:dTbFit} we show the $\delta T_{\rm b}$ from cosmic strings with and without the inclusion of a soft photon heating effect. 
%count -20
it is clear that without the effect we would be in violation of the EDGES bound. The proposed solution therefore represents an important example of a model which produces a RSB at cosmic dawn, but does not violate global $21$ cm observations.

%%%%%%%%%%%%%%%%%%%%%%%%%%%%%%%%%%%%%%%%%%%%%%%%%%
%%%%%%%%%%%%%%%%%%%%%%%%%%%%%%%%%%%%%%%%%%%%%%%%%%

%%%%%%%%%%%%%%%%%%%%%%%%%%%%%%%%%%%%%%%%%%%%%%%%%%%%%%%%%%%%%%%%
\section{\label{sec:level4}Conclusions}
%%%%%%%%%%%%%%%%%%%%%%%%%%%%%%%%%%%%%%%%%%%%%%%%%%%%%%%%%%%%%%%%
In this work, we have studied the low-frequency spectral signatures produced by a network of superconducting string loops. Using \texttt{CosmoTherm}, 
%to perform a grid search of models, 
%count -7 
we found a region in the $G\mu$-$\mathcal{I}$ parameter space that offers a convincing explanation to the observed RSB (see Fig.~\ref{fig:LowFreqFits}). After considering constraints from CMB anisotropies, spectral distortions, the optical depth to reionization, and global $21$ cm experiments, we found a best fit solution to the low-frequency radio data with $G\mu \simeq 6.5 \times 10^{-12}$ and $\mathcal{I} \simeq 2.5 \times 10^6$ GeV. 
This string model is only a marginally worse fit to the data when compared against a completely phenomenological power law ($\Delta \chi^2_{\rm r} = 0.66$), and offers an intriguing avenue for further study.

While we make no claim that the radio synchrotron background is a smoking-gun signal of superconducting cosmic strings, it is nonetheless intriguing that this model offers such a superb description of the data. The detection of cosmic strings would offer some much-needed insight into the symmetry breaking patterns experienced by our universe. Thus, we advocate that this simplistic model should be studied in greater detail to fully elucidate its observational signatures.

Superconducting string models are still in need of many refinements. Of particular importance is the simplifying assumption that the current on loops is both time- and length-independent. Indeed, we know that the current on a loop must be generated dynamically by local magnetic fields and thus is expected to evolve. Using large-scale simulations which study the properties and evolution of $\mathcal{I}$, we could improve our prediction of the RSB from such a network. Additionally, the values of $\Gamma_{\rm g}$ and $\Gamma_{\gamma}$ are inferred from simulations, and possess an uncertainty that may alter our best-fit values slightly.

Furthermore, our likelihood analysis is simplistic in a number of ways. Most importantly, a proper marginalization over radio foregrounds should be included. In addition, the CMB likelihood can be computed in a more accurate way which might reveal some differences. However, we believe that none of these should change the main conclusion drastically.

On the observational frontier we highlight that the cosmic string solution could be distinguished from a pure power-law at $\nu \gtrsim 1$~GHz (see Fig.~\ref{fig:LowFreqFits}). Current data from ARCADE-2 and FIRAS do not have the required sensitivity, but in the near future TMS \citep{Jose2020TMS} is set to improve existing measurements at $10-20$~GHz to a sensitivity of 10~Jy/sr. In addition, future measurements with BISOU \citep{BISOU}, COSMO \citep{Masi2021} and a PIXIE-type experiment \citep{Kogut2011PIXIE, Kogut2016SPIE, Chluba2021Voyage} could yield significantly improved limits at $\nu\simeq 30-1000$~GHz (see Fig.~\ref{fig:LowFreqFits}), providing an avenue for testing the model. The obtained 21 cm global signal furthermore departs significantly from the standard prediction (see Fig.~\ref{fig:dTbFit}), opening another way to study the cosmic string interpretation, e.g., with REACH \citep{REACH2022}. The modified global signal also implies that the 21 cm fluctuations should be significantly altered, identifying a new target for 21 cm cosmology using upcoming experiments such as the SKA \footnote{\url{https://www.skao.int/}} and HERA \cite{HERA2021a,HERA2021b}.

%%%%%%%%%%%%%%%%%%%%%%%%%%%%%%%%%%%%%%%%%%%%%%%%%%%%%%%%%%%%%%%%
\section*{Acknowledgements}
%%%%%%%%%%%%%%%%%%%%%%%%%%%%%%%%%%%%%%%%%%%%%%%%%%%%%%%%%%%%%%%%
\begin{acknowledgments}
%We would like to thank Jack Singal, J.A. Rubino-Martin, Robert Brandenberger, and Richard Battye for helpful comments on the draft.
%
This work was supported by the ERC Consolidator Grant {\it CMBSPEC} (No.~725456).
JC was furthermore supported by the Royal Society as a Royal Society University Research Fellow at the University of Manchester, UK (No.~URF/R/191023).
BC would also like to acknowledge support from an NSERC-PDF.
\end{acknowledgments}

\bibliographystyle{apsrev4-1}
\bibliography{main}% Produces the bibliography via BibTeX.

\end{document}